\documentclass[prd,referee,nopacs,nofootinbib, twocolumn]{revtex4-1}
\usepackage[colorlinks,linkcolor={red},citecolor={blue}]{hyperref}
\usepackage{amsfonts}
\usepackage{amsmath}
\usepackage{eurosym}
\usepackage{amssymb}
\usepackage{graphicx}
\usepackage{color}

\newcommand{\f}[2]{\frac{#1}{#2}}
\def\be{\begin{equation}}
\def\ee{\end{equation}}
\def\bea{\begin{eqnarray}}
\def\eea{\end{eqnarray}}

\begin{document}
	
	\title{Anisotropy in constraint 4D Gauss-Bonnet gravity}

	\author{Shahab Shahidi$^\alpha$}
	\email{s.shahidi@du.ac.ir}
	\author{Nima Khosravi$^\beta,^\gamma$}
	\email{n-khosravi@sbu.ac.ir}

	\affiliation{$^\alpha$School of Physics, Damghan University, Damghan, 41167-36716, Iran.}
	\affiliation{$^\beta$Department of Physics, Shahid Beheshti University, 1983969411,  Tehran, Iran,\\$^\gamma$School of Physics, Institute for Research in Fundamental Sciences (IPM), P.~O.~Box 19395-5531, Tehran, Iran}
	
	\date{\today}
	
	\begin{abstract}
Recently a new 4D Einstein-Gauss-Bonnet theory has been introduced \textbf{[Phys. Rev. Lett. 124 (2020) 081301]} with a serious debate that it does not possess a covariant equation of motion in $4D$. This feature, makes impossible to consider non-symetric space-times in this model, such as anisotropic cosmology. In this note, we will present a new proposal to make this happen, by introducing a Lagrange multiplier to the action which eliminates the higher dimensional term from the equation of motion. The theory has then a covariant $4D$ equation of motion which is useful to study the less symmetric metrics. On top of FRW universe, the constraint theory is equivalent to the original $4D$ Einstein-Gauss-Bonnet gravity. We will then consider the anisotropic cosmology of the model and compare the theory with observational data. We will see that the theory becomes non-conservative and the matter density abundance falls more rapidly at larger redshifts compared to the conservative matter sources.
	\end{abstract}
	
	\maketitle
	
\section{Introduction}
The Einstein's theory of general relativity has witnessed serious observational and theoretical challenges since its birth in 1915 as a way to describe the gravitational interaction. At the beginning, the theory has been challenged from reconciling the common sense with the predictions of the theory and afterward from solar system/local tests to cosmological/large scale experiments. However, one of the most controversial issued about the theory is to correctly describing the late time accelerated expansion of the universe. The first proposal was given by Einstein himself, considering the cosmological constant to the original Einstein-Hilbert action \cite{cosmologicalconstantreview}. This idea together with the inclusion of cold dark matter to elucidate local observations, is now well-known as the $\Lambda$CDM model
\begin{align}
S=\int d^4x\sqrt{-g}(R-2\Lambda)+S_m,
\end{align}
where $\Lambda$ is the cosmological constant, and $S_m$ is the action for all baryonic matter fields plus dark matter.

Nowadays, we believe that the cosmological constant, suffers from theoretical/phenomenological problems \cite{cosmologicalconstantreview}, tempting cosmologists to consider other options, generally known as dark energy models. The simplest possibility of adding a dark energy to the Einstein's theory is to promote the cosmological constant to a dynamical field. This field could be a scalar, vector or in general some higher spin field \cite{scalarvectortheories}. However, one can consider some modifications of the Einstein-Hilbert action itself which could involve higher dimensions \cite{higherdimension}, granting mass to the mass-less graviton \cite{massivegravity}, considering richer geometries \cite{weylcartan}, or changing the gravitational interactions by modifying the Ricci scalar in the Einstein-Hilbert action \cite{fRreview}. All of these ideas are generally known as modified theories of gravity.

Changing Ricci scalar in the action could be done in numerous ways. The simplest way is to substitute the Ricci scalar with an arbitrary function of $R$, resulting in a so-called $f(R)$ theory of gravity \cite{fRreview}. The theory can be easily proven to be equivalent to a specific class of Brans-Dicke theory and hence it is free of Ostrogradski ghost. As a result it can be safely considered as a geometrical candidate for dark energy. Another possibility to generalize the Ricci scalar in the Einstein-Hilbert action is to add some higher order terms constructed from Ricci and Riemann tensors. In general these higher order terms possess extra unhealthy degrees of freedom which will eradicate the validity of the theory. It is well-known that the only terms which cancels the ghost degrees of freedom would be obtained  from the Lovelock prescription \cite{lovelock}. However, Lovelock terms are all vanished in four dimensional space-times and as a result only the Ricci scalar remains as a healthy candidate to describe gravitational interactions. The first non-trivial Lovelock term is the Gauss-Bonnet invariant which is non-zero for $D>4$, where $D$ is the space-time dimension. An interesting idea to put the effects of this term in the theory is to assume that the universe is fundamentally five dimensional. Then we will project the extra dimension to our ordinary four dimensional space-time and the effect of Gauss-Bonnet term will appear in our $4D$ universe \cite{brane}. 

Recently, an idea appears in the literature, presenting a new way to make some non-vanishing effects of the Gauss-Bonnet Lagrangian in $4D$ \cite{4Dfirst}. The idea comes from the fact that the vanishing of the Gauss-Bonnet invariant in $4D$ is due to the presence of an overal factor $D-4$. The author then claimed that if one considers a $D$-dimensional abstract space-time and re-scales the Gauss-Bonnet parameter $\alpha$ to $\alpha/(D-4)$,  one then obtain some non-vanishing contributions of the Gauss-Bonnet invariant in $4D$ by taking the limit $D\rightarrow4$. The consequences of the theory is widely investigated in the literature, exploring its effects in black-holes \cite{blackholes}, wormholes \cite{wormholes}, compact stars \cite{compactstars} and also in cosmology \cite{othercosmology}. Also, many works has been done to generalize the idea to higher/lower dimensions, relations to other theories and also its quantum aspects of the theory \cite{other works}. For example, in \cite{cao}, the authors showed that in order to have a well-defined linearized theory the metric should be locally conformally flat. Also, many works have been done to constraint the new re-scaled Gauss-Bonnet parameter from observational data \cite{constraintonparameter}. In \cite{mota} the authors obtained the so far strongest constraint on the theory parameter, namely, $(\alpha=2.69\pm 11.67)\times 10^{48}eV^{-2}$. 

However, the most important issue related to the $4D$ Einstein-Gauss-Bonnet theory is that despite the fact that the Lagrangian has an overall factor $D-4$, this is not the case for the equation of motion. As have been shown in the literature \cite{comment} and we will also review in the next section, the equation of motion can be decomposed to two separate parts, one has the $D-4$ factor and the other which is proportional to the Weyl tensor, does not has the factor. As a result, for a generic non-symmetric space-time, the second part will make the equation of motion divergent. Also, it has been pointed out that for non-symmetric space-times, there is no canonical way to define a higher dimensional metric, and the theory seems ambiguous for generic space-times and the limiting process is not well-defined. As a result, the theory does not have a covariant $4D$ equation of motion. 

There are some attempts in the literature to resolve these problems, for example to couple the theory to a scalar field \cite{comment}.
In this paper, we put forward a new idea to deal with the above issues. As we have seen above, the main difficulty for having a covariant equation of motion is that the Gauss-Bonnet term has a term without the factor $D-4$. As a result in the limiting process this term causes problems for generic space-times. We will covariantly set this part to zero from the action by imposing a Lagrange multiplier. This will allow us to have a covariant $4D$ equation of motion and the limiting process would be unambiguous. 

In this paper, we will show that the isotropic cosmology of the new model would be identical to the original theory. We will then consider the anisotropic universe in this model. It should be noted that the same process could not be done in the original theory since, as we discussed above, there is an ambiguity in defining the higher dimensional metric. In our new framework, however, there is no need for higher dimensional metric since we have a covariant  4D equation of motion.
\section{The model}
Let us start with the 4D Gauss-Bonnet gravity theory introduced in \cite{4Dfirst}. The action in $D$ dimensions can be written as
\begin{align}\label{act}
S=\int& d^{D}x\sqrt{-g}\bigg[\kappa^2(R-2\Lambda)+\mathcal{L}_m+\beta\,\mathcal{G}\bigg],\
\end{align}
where $\Lambda$ is the cosmological constant, $\mathcal{L}_m$ is the matter Lagrangian and $\mathcal{G}$ is the Gauss-Bonnet term defined as
\begin{align}\label{GB}
\mathcal{G}=R^{\mu\nu\alpha\beta}R_{\mu\nu\alpha\beta}-4R^{\mu\nu}R_{\mu\nu}+R^2.
\end{align}
The 4D Gauss-Bonnet recipe \cite{4Dfirst} is to rescale the Gauss-Bonnet coupling constant $\beta$ to $\alpha/(D-4)$, so that the Gauss-Bonnet term acquire a  non-vanishing part in the limit $D\rightarrow4$. 

The equation of motion of the metric field in $D$ dimensions can be obtained from the action \eqref{act} as
\begin{align}\label{fe}
\kappa^2 &\left(G_{\mu\nu}+\Lambda g_{\mu\nu}\right)+\f{\alpha}{D-4}\mathcal{G}_{\mu\nu}=\f12T_{\mu\nu},
\end{align}
where we have defined the Gauss-Bonnet tensor
\begin{align}
\mathcal{G}_{\mu\nu}&=2R R_{\mu\nu}-4R_{\mu\alpha}R^{\alpha}_\nu-4R_{\mu\alpha\nu\beta}R^{\alpha\beta}\nonumber\\&+2R_{\mu\alpha\beta\sigma}R_{\nu}^{~\alpha\beta\sigma}
-\f12g_{\mu\nu}\mathcal{G}.
\end{align}
Here, $T_{\mu\nu}$ is the energy-momentum tensor defined as
\begin{align}
T_{\mu\nu}=-\frac{2}{\sqrt{-g}}\frac{\delta \left(\sqrt{-g}\mathcal{L}_m\right)}{\delta g^{\mu\nu}}.
\end{align}
In order to have a covariant $4D$ equation of motion, it is needed that the tensor $\mathcal{G}_{\mu\nu}$ has an overall factor $D-4$. In fact it can be easily verified that the trace of the Gauss-Bonnet tensor has the above property and as a result the  equation of motion \eqref{fe} in $D$ dimensions is
\begin{align}
\kappa^2(D-2)R-2D\kappa^2\Lambda=-T-\alpha\mathcal{G}.
\end{align}
However, in general, this is not the case for the tensor $\mathcal{G}_{\mu\nu}$. Specifically, one can decompose the Gauss-Bonnet part $\mathcal{G}_{\mu\nu}$  \cite{comment} as
\begin{align}\label{gg1}
\mathcal{G}_{\mu\nu}=C_{\mu\nu}+(D-4)S_{\mu\nu},
\end{align}
where
\begin{align}\label{cc}
C_{\mu\nu}=2W_{\mu\alpha\beta\gamma}W_\nu^{~\alpha\beta\gamma}-\f12W_{\alpha\beta\gamma\delta}W^{\alpha\beta\gamma\delta}g_{\mu\nu},
\end{align}
is the Lanczos-Bach tensor where     $W_{\alpha\beta\gamma\delta}$ is the Weyl tensor, and
\begin{align}
S_{\mu\nu}&=\f{2}{(D-2)^2}\Big[DR\,R_{\mu\nu}-2(D-2)R_{\mu\alpha\nu\beta}R^{\alpha\beta}\nonumber\\&+(D-1)g_{\mu\nu}\Big(R_{\alpha\beta}R^{\alpha\beta}+\f{D^2-D+2}{4(D-1)^2}R^2\Big)\nonumber\\&-2(D-1)R_{\mu\alpha}R_\nu^{~\alpha}\Big].
\end{align}
It is then obvious that in the limit $D\rightarrow4$ of the equation of motion \eqref{fe}, the term $S_{\mu\nu}$ remains finite and gives us a non-trivial contribution of the Gauss-Bonnet Lagrangian in $4D$. However, the term $C_{\mu\nu}$ in the field equation \eqref{fe} which is proportional to the Weyl tensor does not have a $(D-4)$ factor and is in general non-vanising. As a result the new factor $1/(D-4)$ could not be canceled in this term and the equation of motion becomes infinite. So, there is no well-defined $4D$ covariant equation of motion of the Gauss-Bonnet theory in this prescription. 

It should be noted that due to the Bach-Lanczos theorem, the very expression in \eqref{cc} identically vanishes in $4D$ space-times. So, the vanishing of the Gauss-Bonnet tensor in $4D$ has two steps. One part becomes identically zero and the other part has an overall factor $D-4$ rendering its vanishing in $4D$. 

In the new approach presented above, we start with a higher dimensional metric, where the tensor $C_{\mu\nu}$ is non-vanishing in general, and then tend $D\rightarrow4$. This causes a two-fold problem. Firstly, we are only interested in the $4D$ space-times with $4D$ metric field. But there is no canonical way to define a higher dimensional metric and as a result, taking the limit $D\rightarrow4$ is not unique. This problem gets worth when the $4D$ metric is asymmetric, like Bianchi space-times. Here, there is no preferred way to define higher dimensional metric.

Secondly, the limit $D\rightarrow4$ is not continuous. This can be seen by considering the fact that the tensor  $C_{\mu\nu}$  is identically zero in $4D$ but it is not in higher dimensions. So we got $\f00$ when we naively take the limit $D\rightarrow4$ and as a result we can not obtain a generally convariant $4D$ equation of motion for Gauss-Bonnet theory.

It should be noted that the above problems seems unimportant for symmetric spacetimes, i.e. when one considers FRW cosmology, blackholes, compact stars, even scalar perturbations around FRW space times, since there is a canonical way to define a higher dimensional metric and for that metric the tensor $C_{\mu\nu}$ vanishes. So, there is no degeneracy in taking the limit $D\rightarrow4$ of the field equation \eqref{fe}. However, this does not imply that the theory has a well-defined covariant equation of motion in $4D$.

It should be mentioned that there is another possibility that we set $C_{\mu\nu}=0$ by hand and only consider the $S_{\mu\nu}$ part of the Gauss-Bonnet term. This solves the problems introduced above, but results in the non-conservation of the energy-momentum tensor. More precisely, in the case of vanishing $C_{\mu\nu}$, the covariant divergence of the field equation \eqref{fe} gives
\begin{align}\label{TT1}
\nabla_\alpha T^\alpha_{~\nu}=\alpha\bigg[&\nabla_\nu\left(R_{\alpha\beta}R^{\alpha\beta}-\f16R^2\right)+R_{\nu\alpha}\nabla^\alpha R\nonumber\\&-2R^{\alpha\beta}\nabla_\beta R_{\nu\alpha}+4R_{\nu\alpha\beta\gamma}\nabla^\gamma R^{\alpha\beta}\bigg],
\end{align}
which is non-zero. Here, a note about this equation is in order. As is well-known, the covariant divergence of any covariant theory is conserved due to the Noether theorem. The covariant divergence of all terms in the LHS of equation \eqref{fe} (in $D\neq4$) are identically zero, which implies that the matter energy-momentum tensor becomes conserved. As we have seen in equation \eqref{gg1}, the Guass-Bonnet term could be decomposed into two terms where the covariant divergence of the tensors $S_{\mu\nu}$ and $C_{\mu\nu}$ are not independently zero. In the $4D$ Gauss-Bonnet theory, the tensor $C_{\mu\nu}$ set to be zero by the limiting process. However, since the covariant derivative of the tensor $C_{\mu\nu}$ is not vanishing, the non-conservative nature of the tensor $S_{\mu\nu}$ remains in the theory which is the RHS of equation \eqref{TT1}. In fact, we have used the conservation equation of the metric field equation to obtain equation \eqref{TT1} which is in agreement with Noether's theorem. We should note that if we insist that the matter is conserved in this theory, we have to constrain the tensor $S_{\mu\nu}$ to be independently conserved (which is equation (11) without the LHS). 

In this paper, we put forward the above argument by imposing the vanishing of the tensor $C_{\mu\nu}$ to the action \eqref{act} through a Lagrange multiplier. The new constrained action becomes
\begin{align}\label{newact}
S=\int d^{D}x\sqrt{-g}\bigg(&\kappa^2(R-2\Lambda)+\mathcal{L}_m\nonumber\\&+\f{\alpha}{D-4}\mathcal{G}+\f12\lambda^{\mu\nu}C_{\mu\nu}\bigg),
\end{align}
where the symmetric tensor $\lambda_{\mu\nu}$ is a Lagrange multiplier. The procedure here is the same as \cite{4Dfirst}; after obtaining the field equations, we perform the limit $D\rightarrow4$ to obtain a $4D$ covariant equation of motion.

The variation should be performed with respect to the metric and the Lagrange multiplier $\lambda_{\mu\nu}$. As a result the theory would have two equations of motions for $g_{\mu\nu}$ and $\lambda_{\mu\nu}$ which can be used to determine all the variables. In our case, the variation of the action \eqref{newact} with respect to $\lambda_{\mu\nu}$ gives $C_{\mu\nu}=0$. We then consider the limit $D\rightarrow4$ of this equation. However, $C_{\mu\nu}$ identically vanishes in $4D$ and there is no equation of motion for the Lagrange multiplier in $4D$. As a result in this model, the Lagrange multiplier would become an arbitrary tensor which could be determined by physical considerations. 

The equation of motion of the metric tensor after taking the limit becomes
\begin{align}\label{newfe}
\kappa^2(G_{\mu\nu}+\Lambda g_{\mu\nu})+\alpha s_{\mu\nu}+L_{\mu\nu}=\f12 T_{\mu\nu},
\end{align}
where we have defined
\begin{align}\label{ss1}
s_{\mu\nu}&=\lim_{D\rightarrow4}S_{\mu\nu}\nonumber\\
&=\f12\Big[4R\,R_{\mu\nu}-4R_{\mu\alpha\nu\beta}R^{\alpha\beta}\nonumber\\&+3g_{\mu\nu}\Big(R_{\alpha\beta}R^{\alpha\beta}+\f{7}{18}R^2\Big)-6R_{\mu\alpha}R_\nu^{~\alpha}\Big],
\end{align}
and
\begin{widetext}
\begin{align}\label{ll}
L_{\mu\nu}&=\f14W^2\lambda_{\mu\nu}+\lambda^{\alpha\beta}W_{\mu\alpha}^{~~\gamma\delta}W_{\nu\beta\gamma\delta}+3\lambda^{\beta\gamma}R^{\alpha}_{~(\mu}W_{\nu)\beta\alpha\gamma}-\f12\lambda R^{\alpha\beta}W_{\mu\alpha\nu\beta}-R^{\alpha\beta}\lambda^\gamma_{(\mu}W_{\nu)\alpha\beta\gamma}+W_{\mu\beta\nu\gamma}R^{\alpha(\beta}\lambda^{\gamma)}_{~\alpha}\nonumber\\&
+2\lambda^{\alpha\beta}W_{\alpha\gamma\beta\delta}W_{\mu~~\nu}^{~\gamma\delta}-2\nabla^\beta\nabla^\gamma W_{\alpha\beta\gamma(\mu}\lambda^\alpha_{~\nu)}-\lambda\nabla^\alpha\nabla^\beta W_{\mu\beta\nu\alpha}-\lambda^{\alpha\beta}\big(\Box W_{\mu\alpha\nu\beta}+2\nabla^\gamma\nabla_\beta W_{\alpha(\mu\nu)\gamma}+2\nabla_{(\mu}\nabla^\gamma W_{\nu)\alpha\beta\gamma}\big)\nonumber\\&
-\f13W_{\mu\alpha\nu\beta}(2R+3\Box)\lambda^{\alpha\beta}+W_{\mu\alpha\nu\beta}\big(\nabla^\alpha\nabla_\gamma\lambda^{\beta\gamma}+\nabla^\beta\nabla_\gamma\lambda^{\gamma\alpha}-\nabla^\beta\nabla^\alpha\lambda\big)+2W_{\alpha\beta\gamma(\mu}\big(\nabla_{\nu)}\nabla^\beta\lambda^{\gamma\alpha}-\nabla^\gamma\nabla^\beta\lambda^\alpha_{~\nu)}\big)\nonumber\\&
-\f18\Big[\lambda W^2-8R^{\alpha\beta}\lambda^{\gamma\delta}W_{\alpha\gamma\beta\delta}+8\lambda^{\alpha\beta}\nabla^\delta\nabla^\gamma W_{\alpha\gamma\beta\delta}+8W_{\alpha\gamma\beta\delta}\nabla^\delta\nabla^\gamma\lambda^{\alpha\beta}\Big]g_{\mu\nu},
\end{align}
\end{widetext}
is the contribution from the new Lagrange multiplier term in the action. In the above expression we have defined
\begin{align}
W^2=W_{\alpha\beta\gamma\delta}W^{\alpha\beta\gamma\delta},
\qquad \lambda=\lambda^\mu_{~\mu}.
\end{align}

Some points should be clarified in this step. First, as we have discussed above, we do not have an equation to determine the Lagrange multiplier $\lambda_{\mu\nu}$. Secondly, one can easily verify that the tensor $L_{\mu\nu}$ vanishes in 4D. This is because this tensor is obtained from variation of the Lanczos-Bach tensor in $D$ dimensions and then taking the $D\rightarrow4$. Since this limiting process continues for this term, the tensor $L_{\mu\nu}$ could be seen as a variation of the Lanczos-Bach tensor in $4D$. Since in $4D$ the Lanczos-Bach tensor vanishes identically, the $L_{\mu\nu}$ does not contribute to the above equations of motion. Thirdly, the energy-momentum tensor is not conserved in this model which can be easily proved by taking the covariant divergence of the equation of motion \eqref{newfe}. The result is exactly equation \eqref{TT1}. Here, the non-conservative nature of the energy-momentum tensor is due to the non-minimal coupling between the Lagrange multiplier term and the metric. It should be noted that the implication of matter conservation from Noether's theorem, works if one have a pure (or minimally coupled) gravity theory. However, due to the Lagrange multiplier term, our model is not purely (or minimally coupled) gravity theory. In summary, the non-conservation of energy-momentum tensor is inherited from the tensor $s_{\mu\nu}$ in \eqref{ss1}.

The resulting equation of motion in for $4D$ Gauss-Bonnet theory in four dimensions can be written as
\begin{align}\label{newfe1}
\kappa^2(G_{\mu\nu}&+\Lambda g_{\mu\nu})+\f{\alpha}{2} \Big[4R\,R_{\mu\nu}-4R_{\mu\alpha\nu\beta}R^{\alpha\beta}\nonumber\\&+3g_{\mu\nu}\Big(R_{\alpha\beta}R^{\alpha\beta}-\f{7}{18}R^2\Big)-6R_{\mu\alpha}R_\nu^{~\alpha}\Big]=\f12 T_{\mu\nu},
\end{align}
Using the definition of the Weyl tensor, one can also write
\begin{align}\label{newfe11}
\kappa^2&(G_{\mu\nu}+\Lambda g_{\mu\nu})+\alpha \Bigg[\f23R\,R_{\mu\nu}-R_{\mu\alpha}R_\nu^{~\alpha}\nonumber\\&+\f12\left(R_{\alpha\beta}R^{\alpha\beta}-\f12R^2\right)g_{\mu\nu}-2R^{\alpha\beta}W_{\alpha\mu\beta\nu}\Bigg]=\f12 T_{\mu\nu},
\end{align}
Actually, if one assumed that the Lanczos-Bach tensor where vanishing from the first place, the same result would appear. In our procedure, this by hand cancellation is replaced by a dynamical cancellation through the Lagrangian.

It should be noted that the new term proportional to $\alpha$ in equation \eqref{newfe11} was obtained from quantum gravity point of view in \cite{qggb} where the authors wrote all the terms which are first order in Weyl tensor with the property that its covariant divergence being zero in conformally flat spacetimes. In this paper, our procedure is classical but we have also kept the first Lovelock invariant which corresponds to the first order Weyl tensor. However, since we have considered the most general spacetimes, the energy-momentum tensor is not covariantly conserved. Rewriting the conservation equation \eqref{TT1} in terms of the Weyl tensor, one obtains
\begin{align}
\nabla^\nu T_{\mu\nu}=2\alpha C_{\mu\alpha\beta\nu}\nabla^\nu R^{\alpha\beta}.
\end{align}
The above equation shows that the matter sector is covariantly conserved in any conformally flat spacetimes which is compatible with \cite{qggb}.
\section{Isotropic cosmology}
As an example of a symmetric space-time, let us try the new constrained 4D Gauss-Bonnet theory on the FRW space-time. Assume that the line element is
\begin{align}
ds^2=a(t)^2\eta_{\mu\nu}dx^\mu dx^\nu,
\end{align}
where $a(t)$ is the scale factor and $\eta_{\mu\nu}$ is the Minkowski metric. We consider a perfect fluid which is characterized  by energy density $\rho$ and thermodynamic pressure $p$, with the energy-momentum tensor given by
\begin{align}
T^{\mu\nu}=(\rho+p)u^\mu u^\nu +p g^{\mu\nu}.
\end{align}
The Friedmann and Raychaudhuri equations can be written as
\begin{align}
3\kappa^2H^2&=\f12a^2\rho+\kappa^2\Lambda a^2-\f{3\alpha}{a^2}H^4,
\end{align}
and
\begin{align}
2\kappa^2\dot H&=\f23\kappa^2\Lambda a^2-\f16a^2(\rho+3p)+\f{2\alpha H^2}{a^2}(H^2-2\dot{H}).
\end{align}
The conservation of the energy-momentum tensor can be written as
\begin{align}
\dot\rho&+3H(\rho+p)=0.
\end{align}
\\
\section{Anisotropic cosmology}
In this section, we will consider the cosmological implications of the anisotropic universe in the constraint 4D Gauss-Bonnet theory. The metric is taken as
\begin{align}
ds^2=-dt^2+a_1^2dx^2+a_2^2(dy^2+dz^2),
\end{align}
where $a_1$, $a_2$ are directional time dependent scale factors. We assume that the universe is filled with an isotropic matter content with energy-momentum tensor of the form
\begin{align}
T^\mu_{~\nu}=\textmd{diag}(-\rho,p,p,p),
\end{align}
where $\rho$ and $p$ are energy density and thermodynamic pressures of the fluid respectively.

Let us define the mean Hubble and the anisotropy parameter as
\begin{align}
H_i&=\frac{\dot{a}_i}{a_i},\quad \Delta H_i=H-H_i\quad i=1,2,\nonumber\\
3H&=H_1+2H_2,\nonumber\\
3A&=\left(\f{\Delta H_1}{H}\right)^2+2\left(\f{\Delta H_2}{H}\right)^2.
\end{align}
Also we define the deceleration parameter as
\begin{align}
q=-1+\frac{d}{dt}\left(\f{1}{H}\right).
\end{align}
It should be noted that in the case of isotropic universe with $a_1=a_2$, the anisotropy parameter vanishes and the mean Hubble and deceleration parameters become the standard Hubble and deceleration parameters of isotropic universe. Since the universe becomes isotropic at late times, we will compare the mean Hubble and deceleration parameters of the anisotropic 4D Gauss-Bonnet universe with late time observational data.
\begin{figure*}
	\includegraphics[scale=0.42]{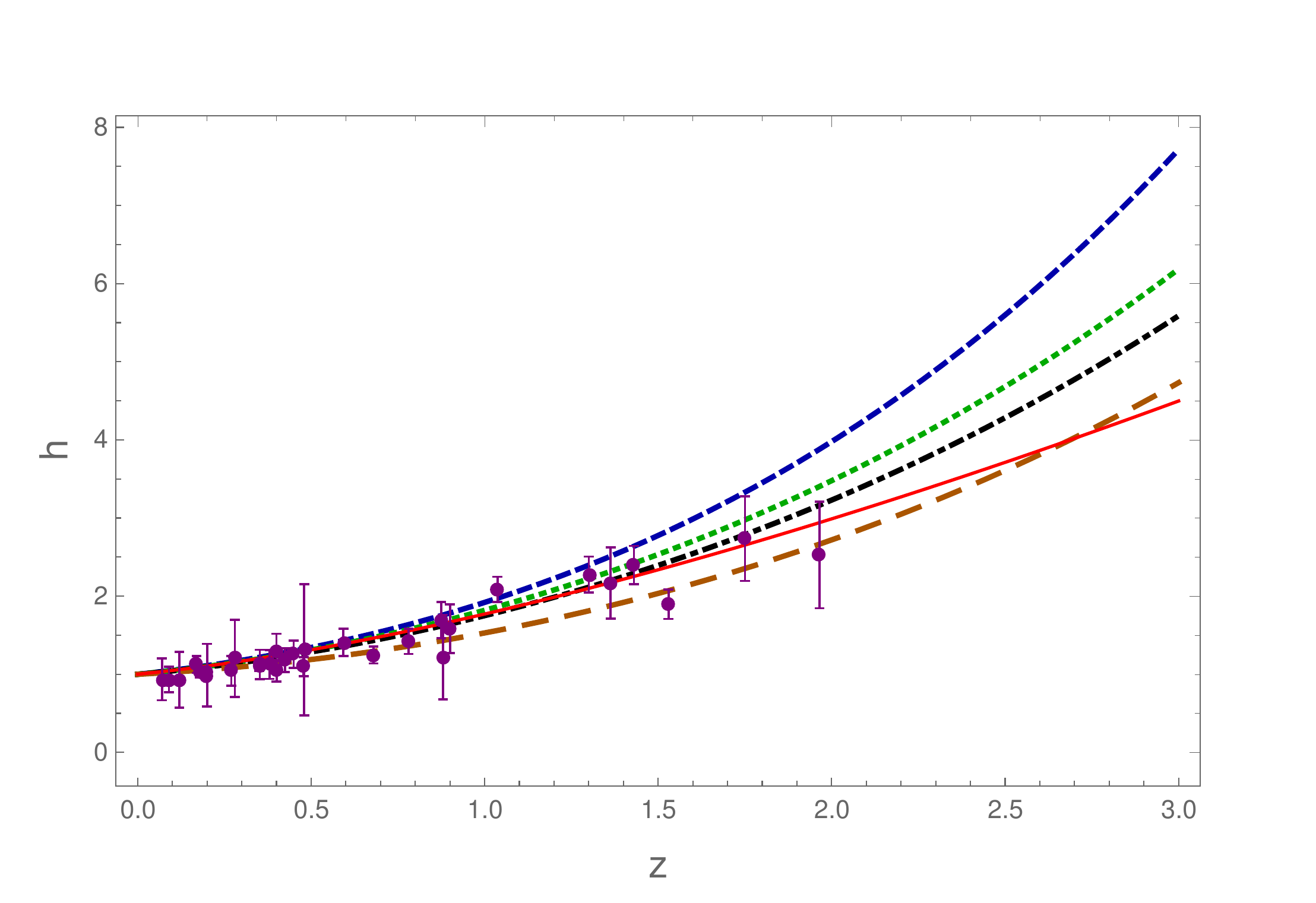}\includegraphics[scale=0.42]{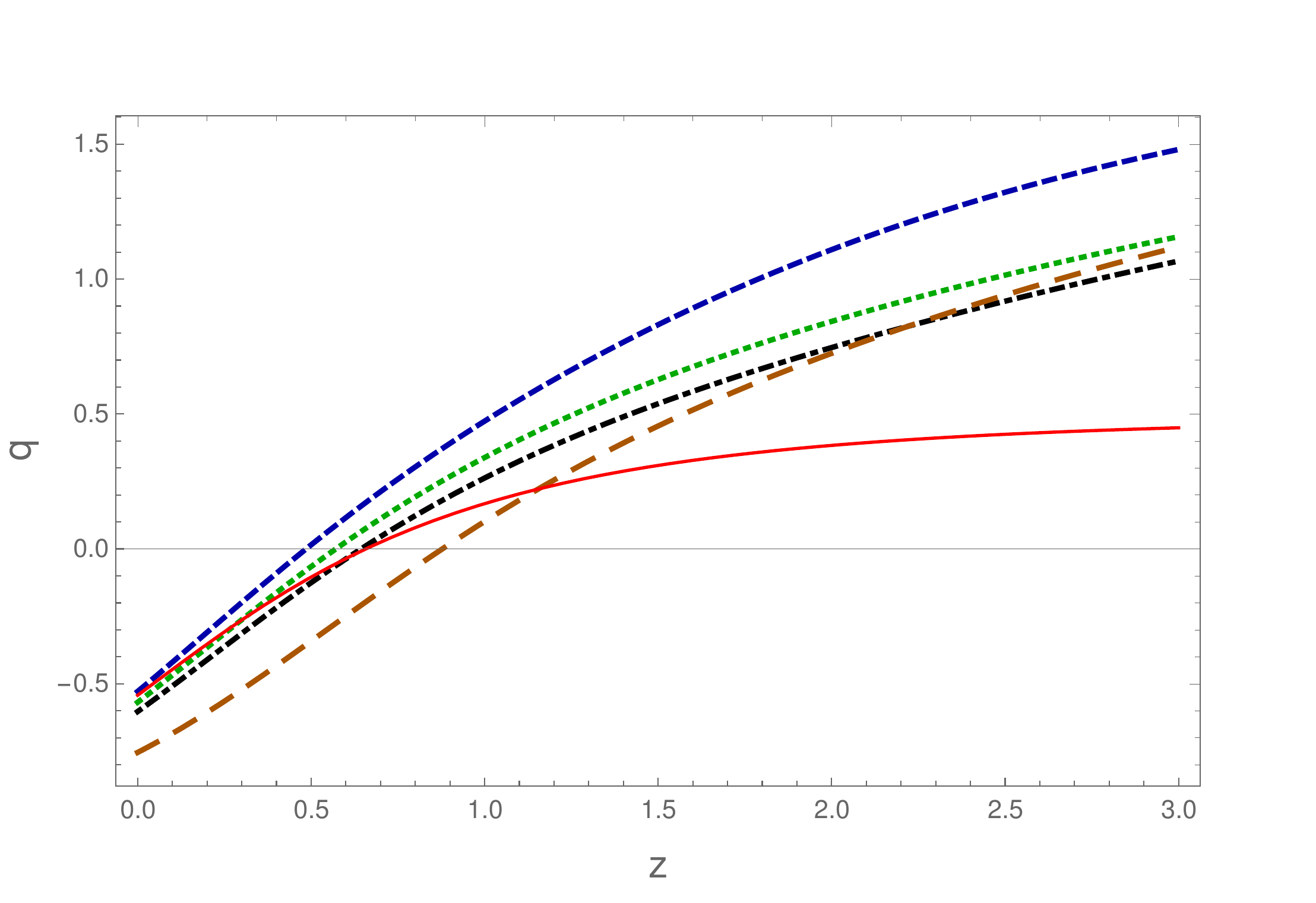}
	\caption{\label{fig1}The evolution of the Hubble and deceleration parameters as a function of redshift $z$ for $\beta=0.001$ (dashed), $0.05$ (dotted), $0.1$ (dot-dashed) and $0.5$ (long dashed). The red solid line represents the evolution in the $\Lambda$CDM model. Also, error bars correspond to the observational data \cite{obshubble}.}
\end{figure*}
The Friedman and Raychaudhuri equations can be written as
\begin{widetext}
\begin{align}
3\kappa^2(1-B^2)H^2&=\kappa^2\Lambda+\f12\rho-3\alpha\Bigg[(1-15B^2+36B^3+3B^4)H^4+2B(1+2B)\dot{B}H^3\nonumber\\ &+2B\dot{B}H\dot{H}+B^2\dot{H}^2+\big(\dot{B}^2+B^2(1+2B)\dot{H}\big)H^2\Bigg],
\end{align}
\begin{align}
\kappa^2\bigg[2(1-B)\dot{H}+3(1-B)^2H^2-2H\dot{B}\bigg]&=-\f12p+\kappa^2\Lambda-\alpha\Bigg[3(5B^4+7B^2-4B+1)H^4-2(2+B)(1-4B)\dot{B}H^3\nonumber\\&+6B\dot{B}H\dot{H}+3B^2\dot{H}^2+\big(3\dot{B}^2+2(4B^3+9B^2-6B+2)\dot{H}\big)H^2\Bigg],
\end{align}
\begin{align}
\kappa^2\bigg[3(1+B+B^2)H^2+H\dot{B}+(2+B)\dot{H}\bigg]&=-\f12p+\kappa^2\Lambda+\alpha\Bigg[3(7B^4+6B^3+8B^2-2B-1)H^4+12B\dot{B}H\dot{H}+6B^2\dot{H}^2\nonumber\\&+2(5B^2+14B-1)\dot{B}H^3+\big(6\dot{B}^2+2(5B^3+18B^2-3B-2)\dot{H}\big)H^2\Bigg].
\end{align}
Also, the conservation equation of the matter content can be written as
\begin{align}
\dot\rho+3H(\rho+p)=&-12\alpha\Bigg(B(1+2B)H^2+\f{d}{dt}(BH)\Bigg)\nonumber\\&\times\Bigg(3B(1-B)(1+2B)H^3+(4+B)\dot{B}H^2+2\dot{B}\dot{H}+B\ddot{H}+\big(\ddot{B}+B(5+B)\dot{H}\big)H\Bigg),
\end{align}
\end{widetext}
where we have defined $B\equiv \sqrt{A/2}$. 
It should be explicitly checked that the tensor $L_{\mu\nu}$ in equation \eqref{ll} vanishes in the case of Bianchi universe. Also, the above conservation equation shows that contrary to the isotropic FRW universe, the matter sector is no longer conserved in this case.

In order to solve the above set of dynamical equations, we assume that the universe is filled with radiation, with energy density $\rho_r$ and pressure $p_r=\rho_r/3$ and also pressure-less dust with energy density $\rho_m$. We then define the following set of dimensionless parameters
\begin{align}
\rho&=\rho_r+\rho_m,\quad H=H_0 h,\quad\beta=\alpha\kappa^2H_0^2,\nonumber\\
\Omega_\Lambda&=\f{\Lambda}{3H_0^2},\quad\bar\rho_i=\frac{\rho_i}{6\kappa^2H_0^2},\quad i=r,m,
\end{align}
where $H_0$ is the current value of the Hubble parameter.

In order to compare the model with observations it is customary to transform the field equations to the redshift coordinates, defined as
\begin{align}
1+z=\f{1}{a}.
\end{align}
We will assume that all the non-conservative sources is handled by dust. As a result the radiation component of the cosmic fluid would be conserved. One can then obtain
\begin{align}
\bar\rho_r=(1+z)^4\Omega_{r0},
\end{align}
where $\Omega_{r0}$ is the current radiation density abundance.

\begin{figure}
	\includegraphics[scale=0.38]{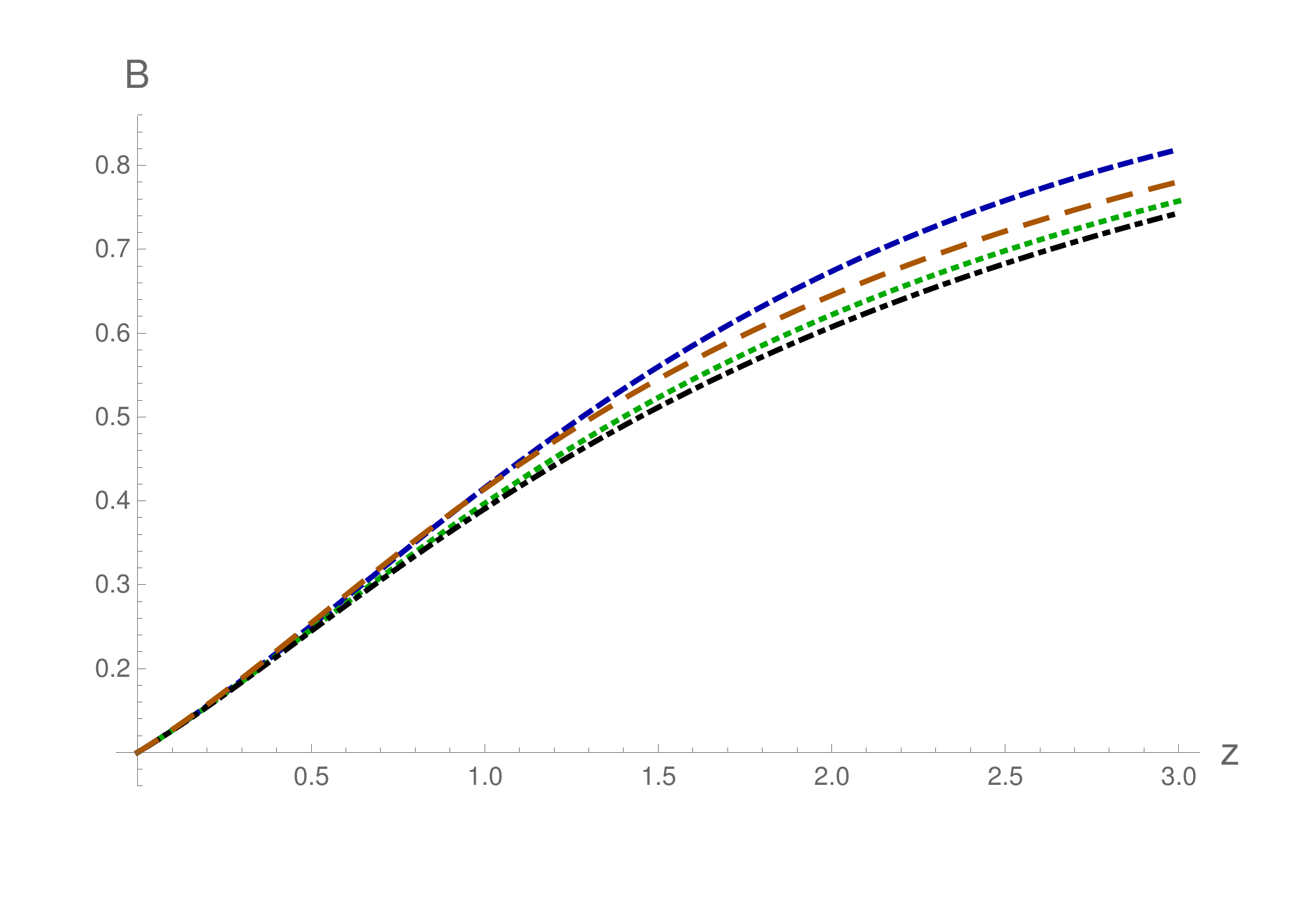}
		\caption{\label{fig2}The evolution of the anisotropy parameter as a function of redshift $z$ for $\beta=0.001$ (dashed), $0.05$ (dotted), $0.1$ (dot-dashed) and $0.5$ (long dashed).}
\end{figure}

\begin{figure}
	\includegraphics[scale=0.38]{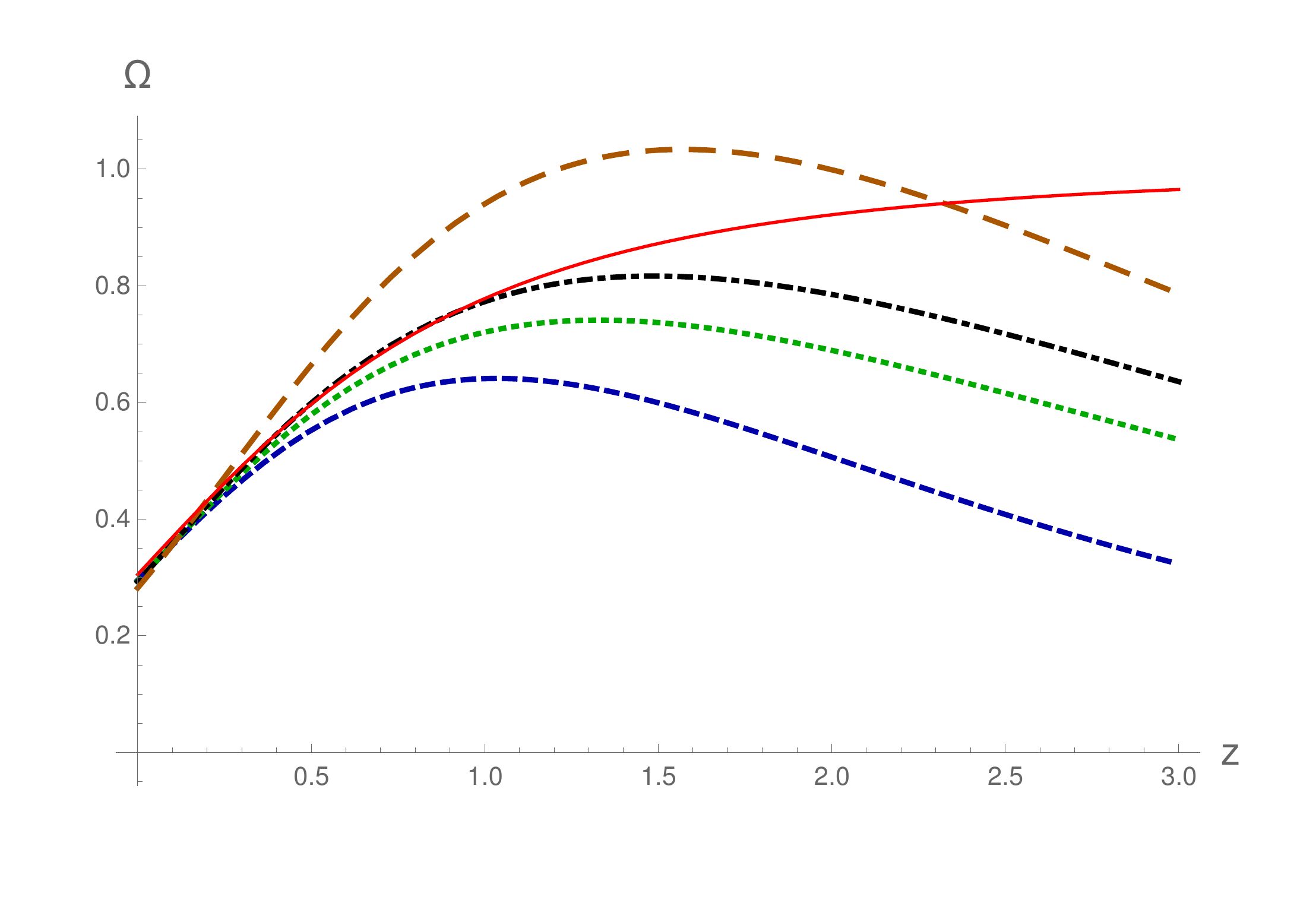}
	\caption{\label{fig4}The evolution of the dust matter density abundance as a function of redshift $z$ for $\beta=0.001$ (dashed), $0.05$ (dotted), $0.1$ (dot-dashed) and $0.5$ (long dashed). The red solid line represents the evolution of the conservative matter source.}
\end{figure}

Considering the Friedman and Raychaudhuri equations at $z=0$ and denoting that by definition $h(z=0)=1$, one can obtain the cosmological constant as
\begin{align}
\Omega_{\Lambda}=1+\beta-\Omega_{r0}-\Omega_{m0},
\end{align}
where $\Omega_{m0}$ is the current value of the dust density abundance. Also, one can obtain the following constraint on the derivatives of the function $B$ and $h$ as
\begin{align}
B^\prime(0)=0,\qquad h^\prime(0)=\f{3\Omega_{m0}+4\Omega_{r0}}{2+4\beta}.
\end{align}
In figure \eqref{fig1}, we have plotted the mean Hubble paramter and also the deceleration parameter as a function of the redshift for different values of $\beta=0.001$ (dashed), $0.05$ (dotted), $0.1$ (dot-dashed) and $0.5$ (long dashed). The red solid line represents the evolution in the $\Lambda$CDM model. The error bars are associated with the observational data on the Hubble parameter \cite{obshubble}. One can see from the figures that the Hubble parameter increases as the parameter $\beta$ decreases. As a result the model predicts smaller universe for smaller values of $\beta$. This can also be seen from the deceleration parameter where smaller values of $\beta$ imply he the universe has more deceleration at larger redshift. In figure \eqref{fig2}, we have plotted the anisotropy parameter $B$ as a function of redshift.
One can see from the figure that the universe becomes isotropic at late times. Also, in figure \eqref{fig4}, we have depicted the dust matter density abundance as a function of $z$. The red solid curve corresponds to the conservative case.
It should be noted that in the 4D-Gauss-Bonnet theory, the matter sector is not conserved. This can be seen from the figure, where the curves fall more rapidly at larger redshifts compare to the conservative case. In summary, we note that the present theory can in principle explain late time observational data. However, more analysis would be needed to determine the viability of the theory as an alternative theory of general relativity.
\section{Conclusions}
In this paper, we have considered a new way to resolve the problem of having an equation of motion in $4D$ Einstein-Gauss-Bonnet gravity by introducing a Lagrange multiplier term to the action. This new term will remove the unwanted part of the Gauss-Bonnet equation of motion which is not proportional to the factor $D-4$ and as a result enabled us to write a covariant equation of motion for the theory. This new term however, make the $4D$ matter field to be non-conservative. In four space-time dimensions, the equation of motion of the Lagrange multiplier satisfied identically and as a result we have an arbitrary tensor field which is not determined in the theory. This signaling a new symmetry which could be unraveled in the future. For symmetric space-times, the new constraint theory is equivalent to the original theory. So, all the works done in the original context could be trivially applied to the new constrained version. The richness of this new constrained 4D Gauss-Bonnet theory is that the field equation is now four dimensional and as a result we could consider less symmetric space-times like Bianchi types or rotating black hole solutions in this context. Importantly, the analysis of perturbations, which needs broken homogeneity and isotropicity, should be done in our framework.  The same analysis could not be done in the original version of the theory because of the ambiguity of defining higher dimensional metrics for these space-times. We have then considered an anisotropic universe describing by the Bianchi type I universe as an example. We have solved the equations numerically and see that the mean Hubble parameter tends to the current Hubble parameter at late times. More precisely, at redshifts smaller than unity, there is no significant deviations between the mean Hubble parameter and the observational data. The anisotropy parameter is also an increasing function of the redshift, predicting that the universe becomes isotropic at late times. Overall, we have shown that the constraint 4D Gauss-Bonnet theory can explain observational data on the Hubble parameter.

It should be noted that in the Bianchi space-time, the matter energy-momentum tensor is no longer conserved. In figure \eqref{fig4}, we have plotted the evolution of the matter density abundance as a function of redshift. It can be seen from the figure that the matter density fall more rapidly at larger redshifts signaling that more matter is transformed to geometry at higher redshifts.

We should note that the non-conservative nature of the matter sector in this model would results in a classical creation of particles. The same property is also considered in other gravitational theories with non-conservative matter energy-momentum tensor \cite{fRT}. This would make a constraint on the parameter $\alpha$. However, we can easily check that the present theory has a Schwarzschild solution with the property that the LHS of equation \eqref{TT1} vanishes. As a result, one concludes that the creation of particles will not happen in this special case. Of course for more general non-Schwarzschild solutions, the creation rate of particles in this theory would be non-zero.

At last, we have to say that more analysis would be needed to fully understand the nature of theory together with is observational constraints.

\section*{acknowledgments}
We would like to thank Zahra Haghani for very useful discussions and comments.

\end{document}